\documentclass[aps,pre,twocolumn,superscriptaddress]{revtex4}

\usepackage{amsfonts,amssymb,amsmath,latexsym,epsfig,wasysym} 
\usepackage[sort&compress]{natbib}

\newcommand{\E}[0]{\mathbb{E}}
\newcommand{\V}[0]{\mathbb{V}}
\newcommand{\N}[0]{\mathcal{N}}
\newcommand{\LN}[0]{\mathcal{LN}}

\begin{document}

\title{Is this scaling nonlinear?}

\author{
J. C. Leit\~ao,  J. M. Miotto, M. Gerlach, and E. G. Altmann}
\affiliation{Max Planck Institute for the Physics of Complex Systems, 01187 Dresden, Germany}


\begin{abstract}
One of the most celebrated findings in complex systems in the last decade is that different indexes
$y$ (e.g., patents) scale nonlinearly with the population~$x$ of the cities
in which they appear, i.e., $y\sim x^\beta, \beta \neq 1$. More recently, the generality of this finding has been questioned in studies using new databases and different definitions of city boundaries. 
In this paper we investigate the existence of nonlinear scaling using a probabilistic framework in which fluctuations are accounted explicitly. In particular, we show that this allows not only to (a) estimate $\beta$ and confidence intervals, but also to (b) quantify the evidence in favor of $\beta \neq 1$ and (c) test the hypothesis that the observations are compatible with the nonlinear scaling. We employ this framework to compare $5$ different models to $15$ different datasets and we find that the answers to points (a)-(c) crucially depend on the fluctuations contained in the data, on how they are modeled, and on the fact that the city sizes are heavy-tailed distributed.
\end{abstract}

\maketitle

\section{Introduction}
The study of statistical and dynamical properties of cities from a complex-systems perspective is increasingly popular~\cite{BattyBook}. A celebrated result is the scaling between a city specific observation $y$ (e.g., the number of patents filed in the city) and the population $x$ of the city as~\cite{Bettencourt2007a}
\begin{equation}\label{eq.scaling}
y = \alpha x^\beta,
\end{equation}
with a non-trivial ($\beta \neq 1$) exponent. Super-linear scaling ($\beta>1$) was
observed when $y$ quantifies creative or economical outputs and indicates that the
concentration of people in large cities leads to an increase in the per-capita production
($y/x$). Sub-linear scaling ($\beta<1$) was observed when $y$ quantifies resource use and
suggests that large cities are more efficient in the per-capita ($y/x$)
consumption.
Since its proposal,  non-linear scaling has been reported in an impressive variety of different aspects of
cities~
\cite{
Bettencourt2010,
Arbesman2011,
Gomez-Lievano2012,
Bettencourt2013a,
Alves2013b,
Nomaler2014,
Oliveira2014}. 
It has also inspired the proposal of different generative processes to
explain its ubiquitous occurrence~\cite{Samaniego2008,
um2009,
Bettencourt2013s,
Pan2013,
Yakubo2014}.
 Scalings similar to the one in Eq.~(\ref{eq.scaling}) appear 
 in physical (e.g., phase transitions) and biological (e.g., allometric scaling) systems suggesting that cities share similarities with these and other complex systems (e.g., fractals).

More recent results cast doubts on the significance of the $\beta \neq 1$ observations \cite{Shalizi2011,Louf2014,Arcaute2015}.
Ref.~\cite{Shalizi2011} agrees that economic outputs are faster than linear in $x$, but
claims that the population $x$ has a limited explanatory factor on the per-capita rate
$y/x$ of cities and  function~(\ref{eq.scaling}) is not better than alternative ones (see Refs.~\cite{Bettencourt2013a, Bettencourt2015} for opposing arguments).
Ref.~\cite{Louf2014} focus on the case of CO2 emissions and show that depending on whether
city boundaries or metropolitan areas are used, the value of $\beta$ changes from
$\beta>1$ to $\beta<1$. This point was carefully analyzed in Ref.~\cite{Arcaute2015} for
different datasets $y$. Through a careful study of different possible choices of city
boundaries, the authors report that the evidence for $\beta \neq 1$ virtually vanishes.  These results ask for a more careful statistical analysis that rigorously quantifies the evidence for $\beta \neq 1$ in different datasets.

\begin{figure*}[!bt]
\includegraphics[width=1.8\columnwidth]{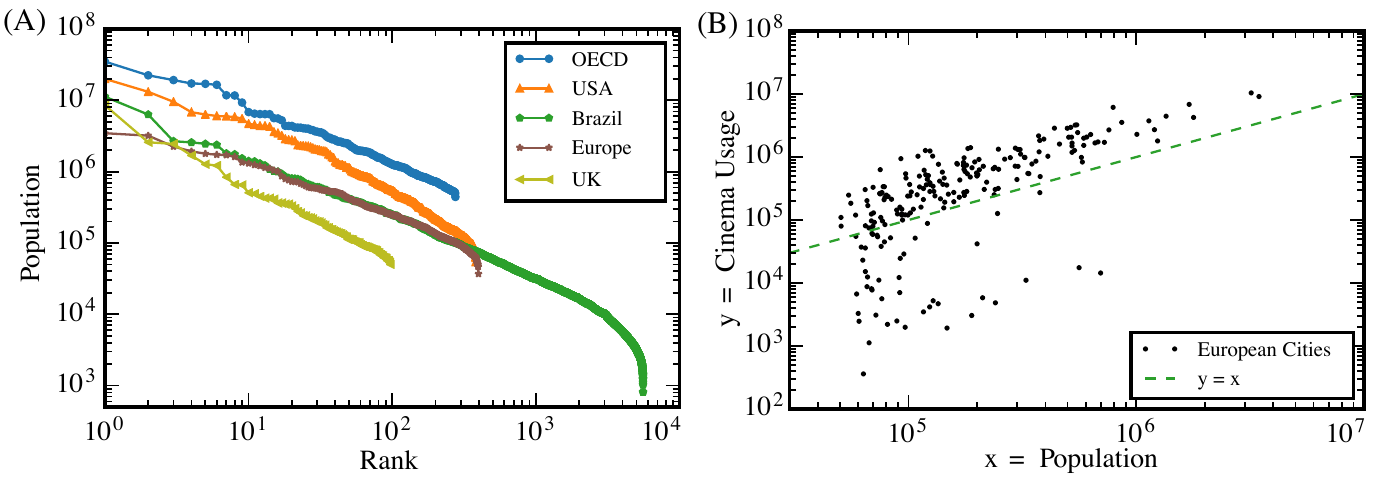}
\caption{Example of the data and its main statistical properties.  (A) The distribution of the population of the cities for the 6
regions considered in this paper. The roughly straight line in this rank-population plot
is in agreement with Zipf's law and shows that, in most cases, the data varies over two
orders of magnitude in population (e.g.,  from $100$ thousands to $10$ million
inhabitants). (B) Example of the dataset analyzed in our work, in which large fluctuations are clearly visible. }
\label{fig_sim}
\end{figure*}

In this paper we propose a statistical framework based on a probabilistic formulation of the scaling law~(\ref{eq.scaling}) that allow us to perform hypothesis testing and model comparison. In particular,  we quantify the evidence in favor of $\beta\neq 1$ comparing (through the Bayesian Information Criterion) models with $\beta\neq1$ to models with $\beta=1$. We apply this approach to $15$ datasets of cities from $5$ regions and find that the conclusions regarding $\beta$ vary dramatically not only depending on the datasets but also on assumptions of the models that go beyond~(\ref{eq.scaling}). We argue that the estimation of $\beta$ is challenging and depend sensitively on the model because of the following two statistical properties of cities:
\begin{itemize}

\item[i] The distribution of city-population has heavy tails (Zipf's law)~\cite{BattyBook,Rybski2013auerbach}.

\item[ii] There are large and heterogeneous fluctuations of $y$ as a function of $x$ (Heteroscedasticity).

\end{itemize}
Points (i) and (ii) are shown, respectively, in panels (A) and (B) of Fig.~\ref{fig_sim}. 

The paper is divided as follows. We start by describing the problem and the datasets we use (in Sec.~\ref{sec.data}) and discussing (in Sec.~\ref{sec.mls})  the limitations of the usual statistical approach based on least-squared fitting in log scale. We then propose a probabilistic formulation together with different statistical models (in Sec.~\ref{sec.models}) and describe (in Sec.~(\ref{sec.results})) how they can be compared to each other and to data.
Finally, we discuss our main findings (in Sec.~\ref{sec.discussion}) and summarize our conclusions (in Sec.~\ref{sec.conclusion}).
\section{Data}

The general problem we are interested in is to test and estimate the parameters of Eq.~(\ref{eq.scaling}) based on observations $(x_i,y_i)$ for $i=1, \cdots, N$ cities, where $x_i$ is the population and $y_i$ is the amount of the quantity of interest in city $i$ (as in Fig.~\ref{fig_sim}B).  The quantities $x_i,y_i$ are estimated within a measurement precision which in principle could also be included in the analysis. However, in most cases this information is not available and only single measurements of $x_i,y_i$ exist. The datasets we choose include a variety of different regions, aggregation methods do define city boundaries, and quantities $y$. It includes data from $5$ different countries and regions: 100 metropolitan areas of the United Kingdom (UK), aggregated as in Ref.~\cite{Arcaute2015}; 381 metropolitan areas of the United States of America (USA), as discussed in Ref.~\cite{Bettencourt2013s}; 459 Urban areas of the USA; 472 large cities of the European Union (EU);  275 large cities from the members of the Organisation for Economic Co-operation and Development (OECD); and 5565 municipalities (administrative units) from Brazil.
For each database, we use indexes of economical activity (weekly income, GDP), innovation (patents filed), transportation (miles traveled, number of train stations), access to culture (number of theaters, number of cinema seats, number of cinema attendances in one year, etc.), and health condition (AIDS infections, death by external causes). 
Further details are presented in the Appendix~\ref{sec.data}.

\section{Limitations of the usual statistical analysis}
\label{sec.mls}

The following three steps summarize the usual approach used to test a non-linear scaling in Eq.~\ref{eq.scaling} (see e.g. Refs.~\cite{
Bettencourt2007a,
um2009,
Bettencourt2010,
Arbesman2011,
Bettencourt2013s,
Louf2014,
Nomaler2014,
Arcaute2015,
Bettencourt2015} for scalings in cities and e.g. Ref.~\cite{Savage2004} for scalings in biology):

\begin{itemize}
\item[1.] The parameters of Eq.~(\ref{eq.scaling}) are chosen based on least-squared
  fitting in log-transformed data $\ln y, \ln x$ , i.e.,  $\alpha,\beta$ are such that $\sum_{i=1}^N (\ln \alpha x_i^\beta - \ln y_i)^2$ is minimized.
\item[2.] The quality of the fitting is quantified by the coefficient of determination $R^2 \equiv 1 - (\sum_i (\ln y_i - \ln \alpha x_i^\beta)^2)/(\sum_i (\ln y_i - \sum_j\ln y_j/N)^2)$. $R^2$ close to $1$ is taken as evidence of the agreement between the fit and the data.
\item[3.] The 95\% confidence interval $[\beta_{\text{min}}, \beta_{\text{max}}]$ around $\beta$ is computed from the sum of the residuals squared and $\beta \not \in [\beta_{\text{min}},\beta_{\text{max}}]$ is taken as an evidence that $\beta \neq 1$.
\end{itemize}

This usual approach is appealing due to its simplicity and ease of numerical implementation. However, it contains the following assumptions and limitations that are usually ignored:

\begin{itemize}
\item[1.] The parameters obtained through least-squared fitting are maximum likelihood estimators if i) the data points are independent and ii) the fluctuations around the mean $\ln y$, $\ln \alpha + \beta \ln x$, are Gaussian distributed in $\ln y$ with a variance  independent of $\ln x$. The value of $\beta$ obtained in the usual approach is meaningful if these assumptions hold.

\item[2.] $R^2$ does not quantify the statistical significance of the model, it quantifies the correlation between data and model (the amount of the variation in the data explained by the model). In particular, $R^2$ close to one is not an evidence that the data is a likely outcome of the model. Below we obtain that  datasets are typically not consistent with the model underlying the usual approach. 

\item[3.] The confidence interval $[\beta_{\text{min}}, \beta_{\text{max}}]$ is a range in which the true value of $\beta$ is expected to be found only if the model holds~\cite{ThulinPhDThesis}. Therefore, in the typical case in which the data is not compatible with the model, one cannot conclude that $\beta \neq 1$ based on the observation that $1 \not \in [\beta_{\text{min}},\beta_{\text{max}}]$. Usually, in this case both $\beta=1$ and $\beta \neq 1$ are incompatible with the data.


\item[4.] A further limitation of the usual approach is that it requires removing the datapoints with $y_i=0$ (because it requires computing $\ln y_i$). This filtering is arbitrary because $y=0$ is usually a valid observation (e.g., cities without any patents filed). 

\end{itemize}

In the study of scaling laws in Biology, the underlying hypothesis and alternatives to the usual  least-squared fitting have been extensively discussed~\cite{Zar1968CalculationData, Warton2006}. 
In city data,  statistical analysis beyond the usual approach were performed in
Refs.~\cite{
Samaniego2008,
Bettencourt2010,
Gomez-Lievano2012,
Alves2013b,
Nomaler2014}. It typically amounts to an analysis  of the residuals $\ln \alpha x_i^\beta -
\ln y_i$, e.g., a (visual) comparison of the residuals of the fit to the Gaussian distribution
predicted by the model underlying the linear fit in  log-log scale.
The controversies regarding a non-linear scaling $\beta \neq 1$ motivate us to search an
alternative statistical framework to test the scaling~(\ref{eq.scaling})  beyond the
usual approach with residual analysis.

\section{Probabilistic models}\label{sec.models}

The statistical analysis we propose is based on the likelihood $\mathcal{L}$ of the data being generated by different models.  Following Ref.~\cite{Gomez-Lievano2012}, we assume that the index $y$ (e.g. number of patents) of a city of size $x$ is a random variable with probability density  $P(y \mid x)$. We interpret Eq.~(\ref{eq.scaling}) as the scaling of the expectation of $y$ with $x$
\begin{equation}\label{eq.scaling2}
\E(y|x) = \alpha x^{\beta} \ \ ,
\end{equation}
where $\E(f(y)|x) \equiv \int f(y) P(y|x) dy$ is computed over the ensemble of cities with fixed $x$.
This relation does not specify the shape of $P(y \mid x)$ , e.g., it does not specify how the fluctuations $\V(y|x)\equiv \E(y^2|x) -\E(y|x)^2$ of $y$ around $\E(y|x)$ scale with $x$. Here we are interested in models $P(y \mid x)$ satisfying
\begin{equation}\label{eq.scaling3}
\V(y|x) = \gamma \E(y|x)^{\delta} \ \ .
\end{equation}
This choice corresponds to Taylor's law~\cite{Taylor}. It is motivated by its ubiquitous
appearance in complex systems~\cite{Eisler}, where typically $\delta \in [1,2]$, and by
previous analysis of city data which reported non-trivial fluctuations
~\cite{Nomaler2014,Hanley2014,Greig2015}.
The fluctuations in our models aim to effectively describe the combination of different effects, such as the  variability in human activity and imprecisions on data gathering. In principle, these effects can be explicitly included in our framework by considering distinct models for each of them.

Below we specify different models $P(y \mid x)$ compatible with Eqs.~(\ref{eq.scaling2},\ref{eq.scaling3}). We consider two classes of models. In the first class, which we call city-models, we {\em a priori} choose a parametric form for $P(y \mid x)$ and we use Eqs.~(\ref{eq.scaling2},\ref{eq.scaling3}) to fix the free parameters. 
In the second class, which we call person-models, we derive  $P(y \mid x)$ from a generative process for the assignment of $y$ to people that is compatible with Eqs.~(\ref{eq.scaling2},\ref{eq.scaling3}).
In both cases, the likelihood $\mathcal{L}$ of the model is written as a function of the data  $\{(x_i,y_i)\}_{i=1,\cdots,N}$  and at most four free parameters ($\alpha, \beta, \gamma$, and $\delta$).

\subsection{City-models}
\label{sec.model.cities}

In this class of models we assume that each data point $y_i$ is an independent realization from the conditional distribution $P(y|x_i)$ and therefore the log-likelihood can be written as
\begin{equation}\label{eq.L}
\ln \mathcal{L} \equiv \ln P(y_1, \cdots, y_N| x_1,\cdots, x_N) =  \sum_{i=1}^{N} \ln P(y_i|x_i).
\end{equation}
In order to explore how the choice of $P(y|x)$ affects the outcome of the statistical analysis, we consider two different continuous distributions (Gaussian and Log-normal)\footnote{This framework allows to use discrete distributions as well.}.

\subsubsection{Gaussian fluctuations}
\label{sec.model.cities.gaussian}
Here we consider that $P(y \mid x)$ is given by a Gaussian distribution with parameters $\mu_{\N}(x)$ and $\sigma_{\N}(x)$:
\begin{equation}
P(y \mid x) = \frac{1}{\sqrt{2\pi}\sigma_{\N}(x)} e^{-\frac{\left(y-\mu_{\N}(x)\right)^2}{2\sigma^2_{\N}(x)}} \ \ .
\end{equation}
The relations~(\ref{eq.scaling2},\ref{eq.scaling3}) are fulfilled choosing the parameters as
\begin{equation}
\begin{alignedat}{2}
\mu_{\N}(x) &= \alpha x^{\beta}\\
\sigma^2_{\N}(x) &= \gamma \left(\alpha x^{\beta}\right)^{\delta} \ \ .
\end{alignedat}
\end{equation}
The log-likelihood~(\ref{eq.L}) is given by
\begin{equation}\label{eq.LN}
\ln \mathcal{L} = \sum_{i=1}^N - \ln (\sigma_{\N}(x_i) \sqrt{2\pi}) - \frac{\left(y_i-\mu_{\N}(x_i) \right)^2}{2\sigma^2_{\N}(x_i)} \ \ .
\end{equation}
This model has $P(y\le 0|x)>0$ and therefore observations with $y_i \le 0$ can be accounted for. For the observables considered here, $y=0$ is a valid observation but $y<0$ is not.

We consider two cases: 

\paragraph{Fixed $\delta = 1$.} This is the typical fluctuation scaling found when $y_i$ is the result of a sum of random variables.~\cite{Eisler}

\paragraph{Free $\delta \in [1,2]$.}  The general functional form that fulfills Eq.(\ref{eq.scaling3}). We exclude $\delta>2$ because in this case the probability $P(y<0 | x)$  of negative values (not feasible for most observables y) remains large for large $x$. 

\subsubsection{Log-normal fluctuations}
\label{sec.model.cities.lognormal}
Here we consider that $P(y \mid x)$ is given by a Log-normal distribution with parameters $\mu_{\LN}(x)$ and $\sigma_{\LN}(x)$:

\begin{equation}
\label{eq.P.lognormal}
P(y \mid x) = \frac{1}{\sqrt{2\pi}\sigma_{\LN}(x)}\frac{1}{y} e^{- \frac{\left(\ln y-\mu_{\LN}(x) \right)^2}{2\sigma^2_{\N}(x)}}.
\end{equation}
The relations~(\ref{eq.scaling2},\ref{eq.scaling3}) are fulfilled choosing the parameters as (see App.~\ref{sec.taylor.logn}):
\begin{equation} \label{eq.ln_parameters}
\begin{alignedat}{2}
\mu_{\LN}(x) &= \ln \alpha + \beta \ln x - \frac{1}{2}\sigma^2_{\LN}(x) \\
\sigma^2_{\LN}(x) &=  \ln \left[ 1 + \gamma \left(\alpha x^{\beta}\right)^{\delta-2} \right].
\end{alignedat}
\end{equation}
The log-likelihood~(\ref{eq.L}) is given by
\begin{equation}\label{eq.LLN}
\ln \mathcal{L} =  \sum_{i=1}^N - \ln (\sigma_{\LN}(x_i) \sqrt{2\pi}) - \ln y_i - \frac{\left( \ln(y_i)-\mu_{\LN}(x_i)\right)^2}{2\sigma^2_{\LN}(x_i)}
\end{equation}
This model has $P(y\le 0|x)=0$ and therefore observations with $y_i \le 0$ cannot be accounted for.

We again consider two cases:
\paragraph{Fixed  $\delta = 2$.} This scaling is obtained when $y_i$ is the product of independent random variables. Furthermore, $\sigma^2_{\LN}(x)$ and the fluctuations of $\ln y$ are independent of $x$ and therefore the maximum likelihood estimation of~$\beta$ coincides with the estimation obtained with minimum least squares for $\ln y$,  as discussed in Sec.~\ref{sec.mls}.

\paragraph{Free $\delta \in [1,3]$.} The general functional form that fulfills Eq.~(\ref{eq.scaling3}).

\subsection{Person-model}
\label{sec.model.people}

The starting point for this class of models is the natural interpretation of
Eq.~(\ref{eq.scaling}) that people's efficiency (or consumption) scale with the size of
the city they are living in.
This motivates us to consider a generative process in which tokens (e.g. a patent,
a dollar of GDP, a mile of road) are produced or consumed by (assigned to) individual
persons, in the same spirit as in Refs.~\cite{Bettencourt2013s,Yakubo2014}.
Specifically, consider $j=1,...,M$ persons living in  $i=1,...,N$ cities, on which the population of the city $i$ is given by $x_i$ such that $\sum_i^N x_i = M$.
Consider also that there is a total of $k=1,...,Y$ tokens that are randomly assigned to the $M$ persons.
A super-linear (sub-linear) scaling suggests that a token is more likely to be assigned to someone living in a more (less) populous city.
In this spirit, we assume that the probability that a token is assigned to person $j$ depends only on the population $x_{(j)}$ of the city where person $j$ lives as
\begin{equation}
p(j) = \frac{x_{(j)}^{\beta-1}}{Z(\beta)},
\end{equation}
where $Z({\beta})$ is the normalization constant, i.e. $Z_\beta=\sum_j^M x_{(j)}^{\beta - 1}$.
For $\beta = 1$, $p(j) = 1/M$ and each person is equally likely to be assigned a token (independently of the population of its city).
The above equation is a microscopic model, and we are now interested in the macroscopic behavior of the city: the probability that a city $i$ gets $y_i$ tokens, given that its population is $x_i$.
Assuming that besides their city, individuals are indistinguishable, the probability $p(i)$ that a token is assigned to a city $i$ is given by a sum of $p(j)$ over persons $j$ on city $i$, which contains exactly $x_i$ terms. Since $x_{(j)} = x_i$ when the person $j$ lives in city $i$, represented by $j\in i$, we obtain
\begin{equation}
  p(i) = \sum_{j\in i} \frac{x_{(j)}^{\beta-1}}{Z(\beta)} = \frac{x_i^{\beta}}{Z(\beta)} \ \ .
\end{equation}
The probability of observing $y_i$ tokens in each city of size $x_i$ is a multinomial distribution 
\begin{equation}\label{eq.multinomial}
  P(y_1, \cdots, y_N | x_1, \cdots, x_N) = Y! \prod_{i=1}^N \frac1{y_i!} \left(\frac{x_i^{\beta}}{Z(\beta)}\right)^{y_i} \ \ .
\end{equation}
Thus, the likelihood can be written as a function of the observed quantities $(x_i,y_i)$ as
\begin{equation}
\begin{alignedat}{2}
\ln \mathcal{L} & \equiv \ln P(y_1, \cdots, y_N | x_1, \cdots, x_N)  \\ 
&= \ln Y! - \sum_{i=1}^N \ln(y_i!) + \sum_{i=1}^N y_i \ln
 \left(\frac{x_i^{\beta}}{Z(\beta)}\right) \ \ .
\end{alignedat}
\end{equation}
The scaling of the average and variance of $y$, i.e. Eqs.~(\ref{eq.scaling2},\ref{eq.scaling3}), are recovered as
\begin{equation}
\begin{alignedat}{2}
 \E(y_i|x_i) &= Y p(i) = \frac{Y}{Z_\beta} x_i^\beta, \\
 \V(y_i|x_i) &= Y p(i) \left[1-p(i)\right] \approx Y p(i) = \E(y_i|x_i) \ \ .
\end{alignedat}
\end{equation}
in which we identify that $\alpha = Y/Z(\beta)$, $\gamma = 1$, and $\delta = 1$. For $y_i \gg 1$, this model coincides with the city-model with normal fluctuations and the latter choice of parameters.
Notice that the fluctuations of this model account only to fluctuations of the assignment, and neglects potential fluctuations of measurement imprecisions.

\section{Results} \label{sec.results}

In this section, we compare the models presented above against our 15 datasets.  In particular, we address the following questions whose answers are summarized in Tab.~\ref{fig_table}:

\begin{table*}[!bt]
\centering\includegraphics[width=1.6\columnwidth]{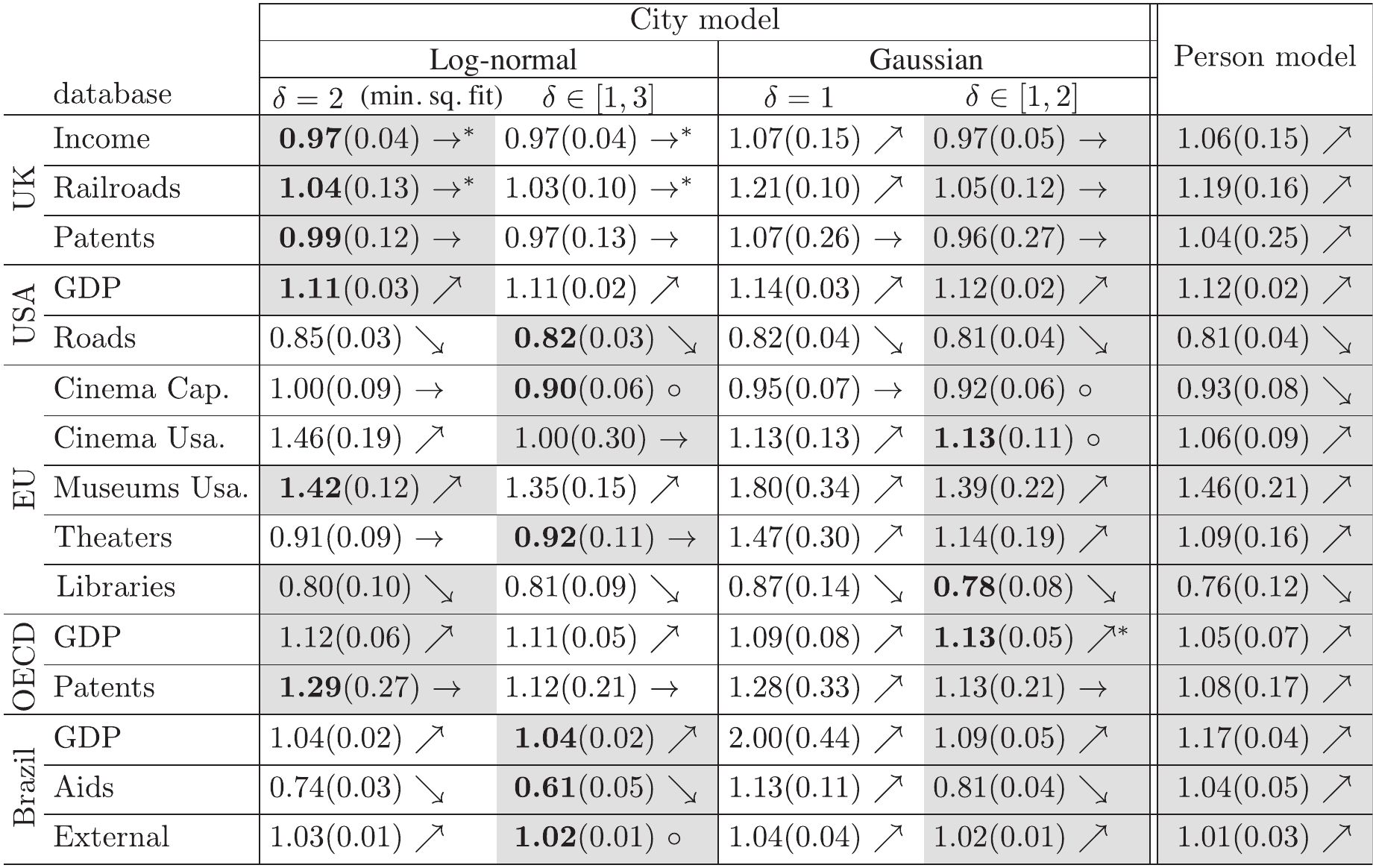}
\caption{Summary of the application of our statistical framework to $15$ different databases and $5$ models. The entries on the tables represent the scaling exponent $\beta$. The value obtained through least-squared fitting in log scale coincides with the value reported in the first column.  The error bars were computed with bootstrap. The $^*$ indicates that the model has a p-value higher than 0.05. If the difference $\Delta BIC$ between the Bayesian Information Criteria (BIC) of each model with the same model with a fixed $\beta = 1$ is below 0, the model is linear ($\rightarrow$), between 0 and 6 is inconclusive ($\circ$), and higher than 6 (strong evidence) is super-linear ($\nearrow$)/sub-linear ($\searrow$). The models were also compared between each other using the respective BICs within the same noise model (gray background has lower BIC) and between all others (bold model has the lowest BIC).
}
\label{fig_table}
\end{table*}

\subsubsection{Hypothesis testing}

\paragraph{\textit{1. What is the estimated value of $\beta$?}}
  For each model we calculate the parameters ($\alpha, \beta,\gamma, \delta$) that maximize $\mathcal{L}$ (see App.~\ref{sec.ML} for details). 
  In Tab.~\ref{fig_table} we report $\beta$.
  
\paragraph{\textit{2. What is the error bar $b$ around the estimated $\beta$?}}
  We estimate $b$ using bootstrapping with replacement (see App.~\ref{sec.error} for details).
 In Tab.~\ref{fig_table} $b$ is shown in parenthesis. The interval $[\beta - b, \beta+b]$ can be interpreted as the $95\%$ confidence interval of $\beta$ when the model is not rejected. Otherwise, it can be interpreted as the robustness of the estimated $\beta$ against fluctuations in the data (cross validation). 

\subsubsection{Model comparison}

\paragraph{\textit{3. Is the data compatible with the model?}}
  We test the hypothesis that the data was generated by the model. Specifically, for each model we compute a p-value that quantifies i) whether the fluctuations in the data are compatible with the expected fluctuations from the model; and ii) whether the residuals are uncorrelated (see App.~\ref{sec.pvalue} for details).
 In case the model is not rejected, i.e. $p$-value $>0.05$, the corresponding entry in Tab.~\ref{fig_table} is marked by the symbol $*$.

\paragraph{{\textit 4. What is the statistical evidence for $\beta \neq 1$?}}
We quantify the evidence for $\beta \neq 1$ by comparing the maximum likelihood $\mathcal{L}$ of each model with the corresponding model where we fix $\beta=1$.
We account for the different number of free parameters (e.g. to avoid overfitting) by using the Bayesian Information Criterion (BIC), $BIC = -2\ln \mathcal{L} + k \ln N$, where $k$ is the number of free parameters and $N$ the number of observations (see App.~\ref{sec.modelselection} for details).
The difference in the $BIC$, $\Delta BIC \equiv BIC_{\beta=1}-BIC_{\beta}$, indicates whether the model with $\beta \neq 1$ provides a sufficiently better description of the data. 
From this we infer that, for i) $\Delta BIC < 0$ the model with fixed $\beta =1$ (linear scaling) is better; ii) $0 \leq \Delta BIC < 6$ the evidence for $\beta \neq 1$ is inconclusive; and iii) $\Delta BIC \geq 6$ the model with $\beta \neq 1$ (non-linear scaling) is better.
In Tab.~\ref{fig_table} these results are indicated by the symbols i) $\rightarrow$ (linear), ii) $\circ$ (inconclusive); or iii) $\searrow$ (sub-linear) or $\nearrow$(super-linear).

\paragraph{{\textit 5. What is the statistical evidence for fluctuation scaling (Taylor's law)?}}
We quantify the evidence for $\delta \neq 1$ ($\delta \neq 2$), i.e. nontrivial scaling in the fluctuations in Eq.~(\ref{eq.scaling3}), in the models of cities with Gaussian (Log-normal) noise.
Within each class, we calculate $\Delta BIC \equiv BIC_{\delta^*}-BIC_{\delta}$, where we compare the $BIC$'s of the model where i) $\delta$ is fixed ($BIC_{\delta^*}$) and ii) where $\delta$ is a free parameter ($BIC_{\delta}$).
In case of $\Delta BIC>0$, the model with $\delta$ as a free parameter (non-trivial fluctuation scaling) provides a better description of the data (see App.~\ref{sec.modelselection} for details).
In Tab.~\ref{fig_table} the entry for the selected model is highlighted with a gray background.

\paragraph{{\textit 6. Which model best describes the data?}}
We calculate the $BIC$ of each of the 5 models (see App.~\ref{sec.modelselection} for details) and select the one with the lowest BIC as the one that best describes the data.
In Tab.~\ref{fig_table} the $\beta$ of the selected model is printed in bold face.

\begin{figure*}[!bt]
\includegraphics[width=1.8\columnwidth]{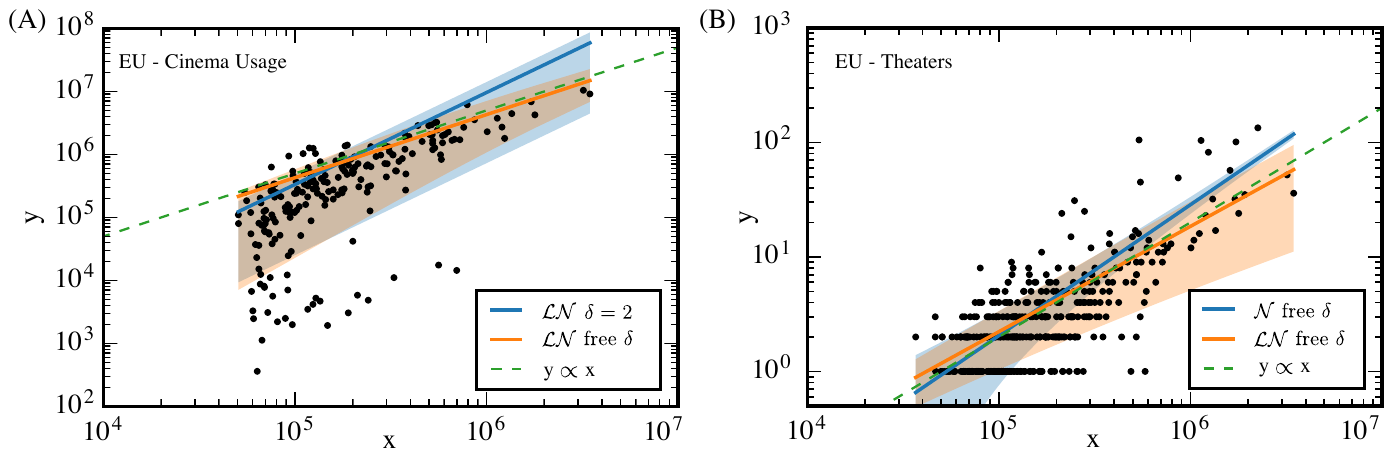}
\caption{Effect of fluctuations on the estimation of $\beta$. (A) In the "EU Cinema Usage" database, the log-normal model with $\delta=2$ yields $\beta=1.46$, while free $\delta$ yields $\beta=1.00$. (B) In the "EU Theaters" database, the log-normal with free $\delta$ yields $\beta=0.92$, a lower value than $\beta=1.14$ obtained in the Gaussian model with free $\delta$. Shaded areas represent the $68$th-percentile ($\pm 1$ standard deviations) of $P(y \mid x)$.}
\label{fig_fluct}
\end{figure*}

\section{Discussion\label{sec.discussion}}

In this section we interpret the outcome of the statistical analysis summarized in 
Tab.~\ref{fig_table}.  We focus on specific findings and their significance to the problem of scaling in cities.

\subsection{Data is almost never compatible with the proposed models}

In almost all cases, the data is not a typical outcome of any of the $5$ proposed models
leading to a rejection of the models (p-value$<$0.05). The only exceptions (marked by an $*$
in the table) are the two log-normal models in UK-Income and UK-Train stations, and the Gaussian model with free $\delta$ for OECD-GDP.
There are several possible reasons for the widespread rejection of the models: fluctuations of the data may differ from the fluctuations $P(y|x)$ of the models (e.g. measurement errors are not correctly accounted for by $P(y|x)$); the observations are not independent (e.g., there are correlations between residuals and city size); different scalings are observed for small and large cities (as discussed in Ref.~\cite{Hanley2016} and Fig.~\ref{fig_city_person} below).

The rejections of the models considered here are a consequence of their strong simplifying hypothesis and show that the development of better models is needed in order to understand the observations and clarify the existence of the non-linear scaling~(\ref{eq.scaling}). It shows also that the estimated confidence interval cannot be used (in the rejected models) to discard a linear scaling $\beta=1$~\cite{ThulinPhDThesis}.
Still, the widespread rejection of models does not imply that the non-linear scaling~(\ref{eq.scaling}) is rejected altogether because it is possible that the data is well described by another (unknown) model consistent with Eq.~(\ref{eq.scaling2}) but different from the ones considered here, e.g., having different fluctuations in $P(y|x)$). These alternative models can have different fluctuation relations or can account for the known (e.g., spatial~\cite{Bettencourt2010}) correlations in the data. In particular, the  generative process underlying the person model could be generalized to account for other effects beyond city-size population (e.g., individuals  could be segmented by income).

Even if most models are rejected, some models can still describe the data better than others (in terms of BIC).  
The conclusions drawn from such {\it model comparison} analysis depends on the used set of models and may change by the introduction of a better model in the future.
Our investigations of scaling laws in cities in the next sections is mostly based on model comparison: we analyze which model and parameters best describe the data, with particular interest in the parameter $\beta$.

\subsection{Different datasets are best described by different models}

There is no single model that best describes all databases (the bold face value in the table appears on different rows). A systematic observation on the $15$ datasets is that the person model and the Gaussian model with fixed $\delta$ are never the best ones. This indicates that the fluctuations in the (large) cities are much larger than predicted by the scaling $\delta=1$ used in both models. 
For the other models, there are databases in which they are the best models: the log-normal with fixed $\delta=2$  is the best model in the three UK cases and for USA GDP; the log-normal model with free $\delta$ is the best model for USA-roads and EU cinema capacity; and the Gaussian model with free $\delta$ is the best for EU-Cinema Usage, OECD-GDP, and EU-Libraries.
The inclusion of the additional parameter $\delta$ in the log-normal model, related to Taylor's law in Eq.~(\ref{eq.scaling3}), is considered beneficial in 8 out of the fifteen approach (shaded gray regions in the two first rows of the table).
Altogether, these results show that the model underlying the usual approach (log-normal with fixed $\delta$) is often not the best model. 

\subsection{The estimated $\beta$ depends on the model}

Models consistent with the average scaling~(\ref{eq.scaling2}), but that have different assumptions regarding the fluctuations, can lead to different estimations of $\beta$. Consider the case of EU-Cinema attendance. The value estimated from the log-normal model with fixed $\delta$ is $\beta=1.46 \pm 0.19$. It coincides with the usual approach (least square fitting) and suggests a super-linear relation between the number of cinema visitors and the population of cities. However, if we allow for a different fluctuation scaling as in the log-normal model with free $\delta$, a model that is preferred according to our BIC test, we obtain that $\beta=1.00 \pm 0.30$, i.e., a linear scaling. Conflicting conclusions are observed also in the EU-Theaters database. The data and fittings for these two cases are shown in Fig.~\ref{fig_fluct}. Visual inspection of the graph can be misleading because of the log-scale and the different density of points, and shows the need for more careful (quantitative) statistical analysis.  Altogether, the variation of $\beta$ across different models shows
that conclusions regarding $\beta$ (e.g., $\beta\neq1$) can not be done independently from the analysis of the fluctuations. Considering also that different models are preferred for different databases (previous point), this confirms the practical importance of going beyond the usual approach (least square fitting) both in terms of methods and models, as proposed in this paper.

\begin{figure}[!bt]
\centering
\includegraphics[width=1\columnwidth]{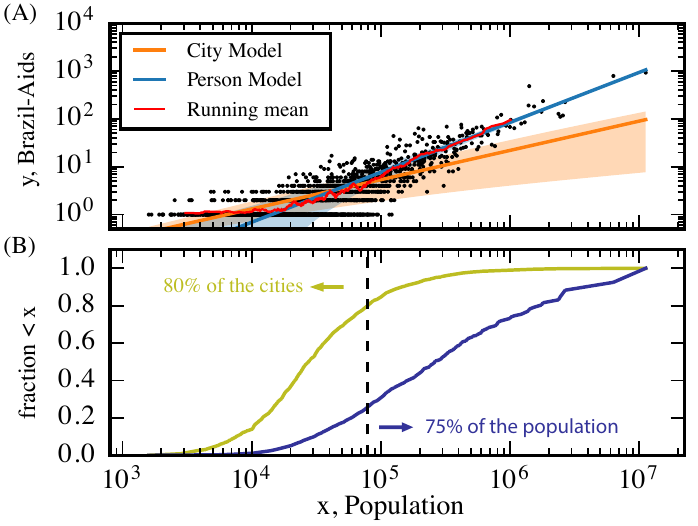}
\caption{Comparison of the Model of Cities and Persons. (A) Scaling of the city model, i.e. Log-normal with free $\delta$, and the person model (solid lines) for the data of Brazil-AIDS (dots). While the city model captures a sub-linear scaling present in small cities $\beta=0.61$, the person model describes the roughly linear scaling $\beta=1.04$ of large cities. Shaded areas represent one standard deviation. The running mean (red line) is the average ($\langle x \rangle$, $\langle y \rangle$) over $50$ datapoints, $\{x,y\}$, in a sliding window over the data ordered in $x$. 
(B) Cumulative distribution of heavy-tailed distribution of city-sizes in terms of cities and persons, i.e. the fraction of i) cities of size $\leq x$ (City Model); and ii) the population in cities of size $\leq x$.
} 
\label{fig_city_person}
\end{figure}

\subsection{Models are dominated either by the small or the large cities}

The variation on the estimation of $\beta$ across the different models can be better understood by analyzing how the city size distribution shown in Fig.~\ref{fig_sim}(A) influences the estimation of $\beta$.
The least-square fitting minimizes the distance between the curve and the points in logarithmic scales ($\ln y$). Therefore, when data is viewed in the usual double logarithmic plot, the best curve will be the one that passes close to most points, i.e., it weights a village as much as a million-size city. The fit will be thus dominated by the large number of small cities. 
The disadvantage of this is that, even if the model describes well most cities, it may fail to describe the behavior of most of the population. Our person's model addresses this issue by giving the same weight to each person, leading to the problem of describing most people but potentially not most cities. 
To see this, consider the example of the 5,565 Brazilian cities. Half of the Brazilian population lives in the $201$ largest cities ($3.6\%$ of cities); yet, 50\% smallest cities account for only $8.2\%$ of the total population. This is a direct consequence of the heavy-tailed distribution of city sizes, which holds in all our databases (see Fig.~\ref{fig_sim}A). 
Our city models with free $\delta$ in Eq.~(\ref{eq.scaling3}) allows cases beyond the least-squared fitting ($\delta=2$) and person's model ($\delta=1$). 
The exponent $\delta$ controls how the variance of $P(y|x)$ grows with $x$. A small variance for large $x$, obtained for small $\delta$, will force the fitted curve (average) to pass close to the points of large cities. 
The weight of the large cities is inversely proportional to $\delta$. 

The general considerations above explain a great extent of the variation of $\beta$ across the models observed in Tab.~\ref{fig_table}. The values obtained for the Gaussian model with $\delta=1$ and the person's model are dominated by large cities, in the log-normal $\delta=2$ case they are dominated by small cities, while for the free $\delta$ models it depend on which best $\delta$ is obtained. In the Brazil AIDS data-set $\delta \geq 2$ and $\beta$ is dominated by the small cities ($\delta=2$ in the Gaussian model, and $\delta=2.79$ in the Log-normal model). Accordingly, the value of $\beta$ for these two models in the second to last row of Tab.~\ref{fig_table} are  $\beta \ll 1$ in agreement with the Log-normal with $\delta=2$ case and in contrast with the Gaussian $\delta=1$ and person model which have $\beta>1$ and are dominated by the large cities. Figure~\ref{fig_city_person} shows the results for this dataset and emphasizes how different models describe different city sizes. The same reasoning explains also the values of $\beta$ of other databases reported in Tab.~\ref{fig_table} (e.g., all UK cases).

In summary, the "weights" each statistical model attributes to cities have an impact on the estimated value of $\beta$ and, in particular, on the visual agreement between the data and the fit in the usual double-logarithmic plots. When the scaling relation~(\ref{eq.scaling2}) holds for all $x$, the difference between the models will not be significant.
However, as we showed in point (a) of this section, data is typically not compatible with
models. In the cases in which $\beta$ varies substantially across models, generalization
beyond the simple scaling~(\ref{eq.scaling})
~\cite{
Bettencourt2013a}
 should be considered in order to account for the $x$ dependence of $\beta$. In this
 case, the heavy-tailed distribution of city sizes leads many models to be dominated
 either by the large amount of small cities or by the few cities containing most of the
 population. This reasoning provides an explanation for why cutoff in minimum city size
 and  aggregation of cities (different city borders)~\cite{Oliveira2014,Louf2014,Arcaute2015}
 influence the estimated $\beta$. All theses procedures have a strong influence on the small cities, which are the dominant ones in the least-square fitting (e.g., aggregation of cities into metropolitan areas reduces the number of small cities). While applying cut-offs for small cities increase the visual agreement between the data and the fit in the log-log plot, this is only justified if the scaling~(\ref{eq.scaling}) is interpreted as being valid only for large cities. The latter interpretation limits the relevance of the scaling which becomes limited to a small fraction of the total cities.

\subsection{Is the scaling nonlinear?}

New answers to this central question emerge from the results of our manuscript (summarized
in Tab.~\ref{fig_table}). In $3$ of the $15$ cases we found models which are reasonably
compatible with the data and we can base our conclusions on these models, i.e., on the
obtained $\beta$ and on the model comparison to the case $\beta=1$ (arrows
$\rightarrow,\uparrow,\searrow$ in the Table). This leads to the conclusion that the
UK-Income and UK-Train stations show linear and OECD-GDP shows superlinear scaling. 
In the remaining 12 cases, conclusions are based solely on model comparison and we feel more confident to give an answer to this question only when the same conclusion is obtained for models with different fluctuations (i.e., we compare the conclusions obtained in the best model with Log-normal and Gaussian fluctuations). We find such an agreement in 8 of the 12 cases so that the scaling: UK-Patents and OECD-Patents are linear; USA-GDP, EU-Museum Usag, and Brazil-GDP are superlinear; USA-Roads, EU-Libraries, and Brazil-AIDS are sublinear. For the remaining $4$ cases our analysis is \textit{inconclusive} on the question of linear or nonlinear scaling. 
Two reasons can lead to this conclusion. The first is that the nonlinear scaling qualitatively changes from $\beta<1$ to $\beta>1$ depending on the assumptions of the fluctuations (e.g. EU N. Theaters).
The second reason is that in one of the best models there is no sufficient statistical evidence for $\beta\neq1$ (marked by a $\circ$ in the Table, EU-Cinema Capacity,Eu-Cinema Usage, and Brazil-External). One interesting case falling in this second reason is EU-Cinema Usage, for which both the log-normal with fixed $\delta$ and the best model (Gaussian with free $\delta$) yield $\beta>1$. We still consider this case inconclusive because the best model, despite showing $\beta=1.13\pm0.11$, only marginally improves ($0<BIC<6$) upon the model with $\beta=1$. In this case, additional data is required in order to increase the statistical evidence in favor of either situation. The possibility of reaching an inconclusive answer shows the advantage of the statistical framework proposed here. In summary, in $15$ datasets we found $4$ linear, $4$  super-linear, and $3$ sub-linear scalings.

\section{Conclusions\label{sec.conclusion}}

In summary, we investigated the existence of non-trivial $\beta\neq1$ scalings in city datasets. We introduced $5$ different models, showed how to compare them and how to estimate $\beta$, and finally tested our methods and models in $15$ different datasets. 
We found that in most cases models are rejected by the data and therefore conclusions can only be based on the comparison between the descriptive power of the different models considered here. Moreover, we found that models which differ only in their assumptions on the fluctuations can lead to different estimations of the scaling exponent $\beta$.  In extreme cases, even the conclusion on whether a city index scales linearly $\beta=1$ or non-linearly $\beta\neq1$ with city population depends on the assumptions on the fluctuation. A further factor contributing to the large variability of $\beta$ is the broad city-size distribution which makes models to be dominated either by small or by large cities. In particular, these results show that the usual approach based on least-square fitting is not sufficient to conclude on the existence of non-linear scaling.

Recent works focused on developing generative models of urban formation that explain
non-linear scalings~
\cite{Samaniego2008,
um2009,
Bettencourt2013s,
Pan2013,
Yakubo2014}.
Our finding that most models are rejected by the data confirms the need for such improved models. The significance of our results on models with different fluctuations is that they show that the estimation of $\beta$ and the development of generative models cannot be done as separate steps. Instead, it is essential to consider the predicted fluctuations not only in the validation of the model but also in the estimation of $\beta$.
Finally, the methods and models used in our paper can be applied to investigate scaling laws beyond cities~\cite{Savage2004,Warton2006}.

\section*{Acknowledgment}

We thank E. Arcaute for kindly sharing the UK databases and D. Rybski and L. Bettencourt for helpful discussions.

\section*{Appendices}
\label{sec.App}

\subsection{Databases}
\label{sec.data}

We used 15 datasets from 5 different databases. In each database (UK, USA, EU, OECD,
Brazil), the same cities $x_i$ were used, and the different datasets are different indexes
$y$. Some of our models cannot consider $y_i\leq 0$. In order to allow for a comparison
across all models,  we ignored $y_i\leq 0$ in all cases and below we report the number $N$
of cases $y_i>0$ in each dataset. 

\begin{itemize}
\item UK: this database corresponds to Fig.~5b of Ref.~\cite{Arcaute2015}, was provided
by the authors of that paper, include the aggregation of population in cities proposed
in that paper, and corresponds to period 2000-2011.
  \begin{itemize}
  \item Income: $N=100$, total income (weekly).

  \item Train stations: $N=97$, number of train stations.

  \item Patents: $N=93$, number of Patents filed in the period.

  \end{itemize}

\item USA: This database corresponds to metropolitan areas of the USA (GDP) and Urban Areas (Roads) in 2013. It was
  constructed from 3 different sources: the population was provided by U.S. Census
  Bureau~\cite{us_census}; the GDP was provided by the U.S. Bureau of Economics Analysis
  of the Department of Commerce~\cite{us_gdp}, and the Miles of roads was provided by the
  U.S. Federal Administration of Highways of the Department of
  Transportation (table HM-71)~\cite{us_roads}. Similar data was used in Ref.~\cite{Bettencourt2013s}.

	\begin{itemize}
	\item GDP: $N=381$, gross domestic product of metropolitan areas.
	\item Roads: $N=459$, length (in miles) of roads of Urban Areas.
	\end{itemize}

\item EU: This database is provided by Eurostat~\cite{eurostat}. It contains population
  and different indexes related to culture in European cities in the year of 2011.

	\begin{itemize}
	\item Cinema Capacity: $N=418$, total number of seats of cinemas.
	\item Cinema Usage: $N=221$, attendance of cinemas in the year.
	\item Museums Usage: $N=443$, attendance of museums in the year.
	\item Theaters: $N=398$, number of theaters.
	\item Libraries: $N=597$, number of public libraries.
	\end{itemize}

\item OECD: This database contains indexes of cities from the Organisation for Economic Co-operation and Development in the years 2000-2012~\cite{oecd}.

	\begin{itemize}
	\item GDP: $N=275$, gross domestic product in 2010.
	\item Patents: $N=218$, number of patents filed in  2008.
	\end{itemize}

\item Brazil: This database contains different indexes of all municipalities of
  Brazil. The data is from the year 2010 and is provided by Brazil's Health
  Ministery~\cite{brdata} (population corresponds to census data).

	\begin{itemize}
	\item GDP: $N=5565$, gross domestic product.
	\item AIDS: $N=1812$, number of deaths by AIDS.
	\item External: $N=5286$, number of deaths by external causes.
	\end{itemize}

\end{itemize}

All the above databases are provided in Ref.~\cite{codeURL}.

\subsection{Taylor's law in log-normal}
\label{sec.taylor.logn}
Here we express the parameters of the log-normal distribution, $\mu_{\LN}(x)$ and $\sigma_{\LN}^2(x)$, as a function of the parameters of the scaling laws
\begin{equation}
\E(y|x) = \alpha x^\beta \tag{\ref{eq.scaling2}},
\end{equation}
\begin{equation}
\V(y|x) = \gamma \E(x)^\delta \tag{\ref{eq.scaling3}} \\,
\end{equation}
$\alpha, \beta, \gamma$ and $\delta$. Noting that the expectation and the variance of the log-normal distribution, Eq.~(\ref{eq.P.lognormal}), are given by
\begin{equation}
\E(y|x) = e^{\mu_{\LN}(x) + \sigma_{\LN}^2(x)/2},
\end{equation}
\begin{equation}
\V(y|x) = (e^{\sigma_{\LN}^2(x)} - 1) \E(y|x)^2 \\.
\end{equation}
we find a unique solution for $\mu_{\LN}(x)$ and $\sigma_{\LN}^2(x)$ by comparing with Eqs.~(\ref{eq.scaling2},\ref{eq.scaling3}):
\begin{equation} \tag{\ref{eq.ln_parameters}}
\begin{alignedat}{2}
\mu_{\LN}(x) &= \ln \alpha + \beta \ln x - \frac{1}{2}\sigma^2_{\LN}(x), \\
\sigma^2_{\LN}(x) &=  \ln \left[ 1 + \gamma \left(\alpha x^{\beta}\right)^{\delta-2} \right]. \\
\end{alignedat}
\end{equation}

\subsection{Maximization of the likelihood}
\label{sec.ML}
The maximization of the likelihood is performed by minimizing minus the log likelihood, using the algorithm "L-BFGS-B"~\cite{Byrd1995}, whose implementation can be found on the Python package scipy~\cite{scipy}, and the details can be found in Ref.~\cite{codeURL}.
Given that the minimization algorithm can converge in a local minimum, our procedure repeats the optimization 512 times, each with random initial parameters; then, we select the among these local minima the lowest, the global minimum. 
We confirmed that increasing from 256 to 512 samples did not change the computed minimum, a confirmation that the algorithm found the global one.

\subsection{Computation of the error estimates}
\label{sec.error}
The error estimates were computed using bootstrap~\cite{Hastie2009}. The method consists in sampling $N$ pairs $(x_i,y_i)$ with replacement from the set of $N$ available data points, and repeat the maximization procedure outlined in the previous section for each set. This procedure (sampling + maximization) was repeated $100$ times for each combination (model, dataset) and the error estimates were computed as the standard deviation of the distances from the measured parameters to the estimated parameter from the true data-set. We confirmed that the bootstrap error estimates for the Log-normal fixed-$\delta$ case are within 1\% equal to the values of the least square fit.

\subsection{Computation of the p-value}
\label{sec.pvalue}
The computation of the p-value was done by defining a statistic that tests the hypothesis used in each model;
in the case of the log-normal and normal models these are: a) data is independent; b) the data is compatible with the model. 
We used a statistic based on the D'Agostino $K^2$ test~\cite{Dagostino1986} (over $\ln y$ or $y$ respectively), that computes the deviations from $0$ of the empirical kurtosis and skewness; the test consist in comparing it with the fluctuations expected from a finite-size sample from the (null) model. 
In detail, the we compute two statistics, $Z_s$ and $Z_k$ for the kurtosis and skewness respectively.
Each of them has a $\chi^2_1$ distribution under the null, so the sum, $K^2 = Z_s^2 + Z_k^2$ is has a $\chi^2_2$ distribution (with 2 degrees of freedom).
Because this test does not test independence of the samples, we include in the test statistic the Spearman's rank-correlation~\cite{Kendall1970} of the residuals of the fit, $Z_S$ (also distributed with as a $\chi_1^2$) because if the residuals are correlated, the data is not independent.
The p-value is thus computed by measuring how extreme $K^2 = Z_s^2 + Z_k^2 + Z_S^2$ is in the $\chi^2_3$ distribution (with 3 degrees of freedom).
The implementation of this is available in the supplementary information \cite{codeURL}.

In the population model the calculation of the p-value must be different, because the variance is not being left as a free parameter, so and we take a more \emph{classical} approach.
The $p$-value is computed by measuring how extreme is the difference between the data and its fit with respect to the difference between a sample from the model and its fit. 
In practice, we use a $\chi^2$ statistic to measure the distance between to sets of points $\{y_i\}_i$ (data) and $\{m_i\}_i$ (the model), $\chi^2 = \sum_i (y_i-m_i)^2/y_i$. Then we generate from the model 200 different samples. For each of these samples we compute the $\chi^2$ between the sample values and their fits. Finally, we compute the $p$-value as the fraction of samples whose $\chi^2$ is bigger than the one that belongs to the real data.
Notice that this statistic is not taking into account independence of the residuals (if we consider the multinomial distribution as the null model, they should not be independent) or normality in the strict sense, so this test is more permissive than the previous.

\subsection{Model comparison using Bayesian Information Criterion}
\label{sec.modelselection}

We compare two models $m=1,2$ by calculating the Bayesian Information criterion (BIC)~\cite{Schwarz1978}, $BIC_m \equiv -2 \ln \mathcal{L}_m + k_m \ln N$, where $N$ is the number of data points (observations), $\mathcal{L}_m$ is the maximum likelihood of the model, $k_m$ is the number of estimated (free) parameters of the model.
In this approach, the model with a lower value for the BIC gives a better description of the data.

We can quantify how much better one model compares to the other by looking at the Bayes' factor~\cite{Kass1995}, $B_{12} = P(\mathrm{data}\mid m=1)/P(\mathrm{data}\mid m=2)$, where $P(\mathrm{data}\mid m)$ is the evidence for model $m$, i.e. the probability of the data given the model.
It can be shown~\cite{Hastie2009} that this quantity can be approximated by
\begin{equation}
B_{12}  \approx e^{1/2 \Delta BIC}
\end{equation}
where $\Delta BIC \equiv BIC_2 - BIC_1$ is the difference of the respective $BIC$'s.
Thus, if $BIC_1<BIC_2$, it follows that $B_{12}>1$, i.e. that model $1$ provides a better description of the data than model $2$.
Regarding the decision about nonlinear scaling, i.e. $\beta \neq 1$, we require that $\Delta BIC \equiv BIC_{\beta=1} - BIC_{\beta} \geq 6$ (see main text), in line with Ref.~\cite{Kass1995}, where it is suggested that this implies strong or very strong evidence for a model with $\beta \neq 1$. This corresponds to $B_{12}\geq e^{3}\approx 20.08$, i.e. it is at least $20$ times more likely that the data comes from a model with $\beta \neq 1$.



\begin{thebibliography}{10}

\bibitem{BattyBook}
Michael Batty.
 {\em The new science of cities}.
 Mit Press, 2013.

\bibitem{Bettencourt2007a}
Lu{\'{i}}s M~a Bettencourt, Jos{\'{e}} Lobo, Dirk Helbing, C.~Kuhnert, and
  Geoffrey~B West.
 Growth, innovation, scaling, and the pace of life in cities.
 {\em Proceedings of the National Academy of Sciences},
  104(17):7301--7306, 4 2007.

\bibitem{Bettencourt2010}
Lu{\'{i}}s M~a Bettencourt, Jos{\'{e}} Lobo, Deborah Strumsky, and Geoffrey~B.
  West.
 Urban scaling and its deviations: Revealing the structure of wealth,
  innovation and crime across cities.
 {\em PLoS ONE}, 5(11):e13541, 11 2010.

\bibitem{Arbesman2011}
Samuel Arbesman and Nicholas~A. Christakis.
 Scaling of prosocial behavior in cities.
 {\em Physica A: Statistical Mechanics and its Applications},
  390(11):2155 -- 2159, 2011.

\bibitem{Gomez-Lievano2012}
Andres Gomez-Lievano, Hyejin Youn, and Lu{\'{i}}s M~a Bettencourt.
 The statistics of urban scaling and their connection to zipf's law.
 {\em PloS one}, 7(7):e40393, 1 2012.

\bibitem{Bettencourt2013a}
Luis M.~a. Bettencourt, Jose Lobo, and Hyejin Youn.
 The hypothesis of urban scaling: formalization, implications and
  challenges.
 {\em arXiv:1301.5919}, 2013.
 
 \bibitem{Bettencourt2015}
Luis M.~a. Bettencourt, and Jose Lobo.
 Urban Scaling in Europe.
 {\em arXiv:1510.00902}, 2015.

\bibitem{Alves2013b}
Luiz G.~A. Alves, Haroldo~V. Ribeiro, Ervin~K. Lenzi, and Renio~S. Mendes.
 Distance to the scaling law: A useful approach for unveiling
  relationships between crime and urban metrics.
 {\em PLoS ONE}, 8(8):e69580, 08 2013.

\bibitem{Nomaler2014}
\"Onder Nomaler, Koen Frenken, and Gaston Heimeriks.
 On scaling of scientific knowledge production in u.s. metropolitan
  areas.
 {\em PLoS ONE}, 9(10):e110805, 10 2014.

\bibitem{Oliveira2014}
Erneson~A Oliveira, Jos{\'e}~S Andrade~Jr, and Hern{\'a}n~A Makse.
 Large cities are less green.
 {\em Scientific reports}, 4:4235, 2014.

\bibitem{Samaniego2008}
Horacio Samaniego and Melanie~E Moses.
 Cities as organisms: Allometric scaling of urban road networks.
 {\em Journal of Transport and Land use}, 1(1):21, 2008.

\bibitem{um2009}
Jaegon Um, Seung-Woo Son, Sung-Ik Lee, Hawoong Jeong, and Beom~Jun Kim.
 Scaling laws between population and facility densities.
 {\em Proceedings of the National Academy of Sciences},
  106(34):14236--14240, 2009.

\bibitem{Bettencourt2013s}
Lu{\'\i}s~MA Bettencourt.
 The origins of scaling in cities.
 {\em science}, 340(6139):1438--1441, 2013.

\bibitem{Pan2013}
Wei Pan, Gourab Ghoshal, Coco Krumme, Manuel Cebrian, and Alex Pentland.
 Urban characteristics attributable to density-driven tie formation.
 {\em Nature communications}, 4:1961, 2013.

\bibitem{Yakubo2014}
K.~Yakubo, Y.~Saijo, and D.~Koro\ifmmode~\check{s}\else \v{s}\fi{}ak.
 Superlinear and sublinear urban scaling in geographical networks
  modeling cities.
 {\em Phys. Rev. E}, 90:022803, Aug 2014.

\bibitem{Shalizi2011}
Cosma~Rohilla Shalizi.
 Scaling and hierarchy in urban economies.
 {\em arXiv:1102.4101}, I(1):15, 2 2011.

\bibitem{Louf2014}
R\'emi Louf and Marc Barthelemy.
 Scaling: lost in the smog.
 {\em Environment and Planning B: Planning and Design},
  41(5):767--769, 10 2014.

\bibitem{Arcaute2015}
Elsa Arcaute, Erez Hatna, Peter Ferguson, Hyejin Youn, Anders Johansson, and
  Michael Batty.
 Constructing cities , deconstructing scaling laws.
 {\em Journal of The Royal Society Interface}, (i):3--6, 2015.

\bibitem{Rybski2013auerbach}
Diego Rybski.
 Auerbach's legacy.
 {\em Environment and Planning A}, 45(6):1266--1268, 2013.

\bibitem{Savage2004}
V.~M. Savage, J.~F. Gillooly, W.~H. Woodruff, G.~B. West, A.~P. Allen, B.~J.
  Enquist, and J.~H. Brown.
 {The predominance of quarter-power scaling in biology}.
 {\em Functional Ecology}, 18(2):257--282, apr 2004.

\bibitem{ThulinPhDThesis}
M.~Thulin.
 {\em On Confidence Intervals and Two-Sided Hypothesis Testing}.
 PhD thesis, Uppsala University, 2014.

\bibitem{Zar1968CalculationData}
Jerrold~H. Zar.
 Calculation and miscalculation of the allometric equation as a model
  in biological data.
 {\em BioScience}, 18(12):1118--1120, 1968.

\bibitem{Warton2006}
David~I Warton, Ian~J Wright, Daniel~S Falster, and Mark Westoby.
 Bivariate line-fitting methods for allometry.
 {\em Biological Reviews}, 81(02):259--291, 2006.

\bibitem{Taylor}
L.~R. Taylor.
 {Aggregation, Variance and the Mean}.
 {\em Nature}, 189(4766):732--735, 1961.

\bibitem{Eisler}
Zolt{\'{a}}n Eisler, Imre Bartos, and J{\'{a}}nos Kert{\'{e}}sz.
 {Fluctuation scaling in complex systems: Taylor's law and beyond}.
 {\em Advances in Physics}, 57(1):89--142, 2008.

\bibitem{Hanley2014}
Quentin~S. Hanley, Suniya Khatun, Amal Yosef, and Rachel-May Dyer.
 Fluctuation scaling, taylor's law, and crime.
 {\em PLoS ONE}, 9(10):e109004, 10 2014.

\bibitem{Hanley2016}
Quentin~S Hanley, Dan Lewis, and Haroldo~V Ribeiro, \emph{{Rural to Urban
  Population Density Scaling of Crime and Property Transactions in English and
  Welsh Parliamentary Constituencies.}}, PloS one \textbf{11} (2016), no.~2,
  e0149546.

\bibitem{Greig2015}
Alastair Greig, John Dewhurst, and Malcolm Horner.
 An application of taylor's power law to measure overdispersion of the
  unemployed in english labor markets.
 {\em Geographical Analysis}, 47(2):121--133, 2015.

\bibitem{us_census}
U.S.~Census Bureau.\\
 www.census.gov/popest/data/metro/totals/2014/.
 November 2014.

\bibitem{us_gdp}
U.S.~Bureau of~Economics~Analysis.
 www.bea.gov/itable/index\_regional.cfm.
 November 2015.

\bibitem{us_roads}
U.S.~Department of~Transportation.
 www.fhwa.dot.gov/policyinformation/statistics/2013/
 November 2015.

\bibitem{eurostat}
Eurostat.\\
 http://ec.europa.eu/eurostat/web/cities/data/database
 November 2015.

\bibitem{oecd}
OECD.
 http://dx.doi.org/10.1787/data-00531-en.
 November 2015.

\bibitem{brdata}
Brazilian~Health Ministry.
 July 2015.

\bibitem{codeURL}
The data and code used to obtain all results in this paper are available online at http://dx.doi.org/10.5281/zenodo.49367 .

\bibitem{Byrd1995}
Richard~H. Byrd, Peihuang Lu, Jorge Nocedal, and Ciyou Zhu.
 {A Limited Memory Algorithm for Bound Constrained Optimization}.
 {\em SIAM Journal on Scientific Computing}, 16:1190--1208, 1995.

\bibitem{scipy}
E.~Jones, T.~Oliphant, P.~Peterson, et~al.
 {SciPy}: Open source scientific tools for {Python}, 2001--.
 {\ttfamily \url{http://www.scipy.org}}.

\bibitem{Hastie2009}
Trevor Hastie, Robert Tibshirani, and Jerome Friedman.
 {\em {The Elements of Statistical Learning}}.
 Springer Series in Statistics. Springer New York, New York, NY, 2nd
  edition, 2009.

\bibitem{Dagostino1986}
R.B. D'Agostino.
 {\em {Goodness-of-Fit-Techniques}}.
 Marcel Dekker, New York, 1986.

\bibitem{Kendall1970}
M.~G. Kendall.
 {\em {Rank Correlation Methods}}.
 Griffin, London, 4th edition, 1970.

\bibitem{Schwarz1978}
Gideon Schwarz.
 {Estimating the dimension of a model}.
 {\em The Annals of Statistics}, 6(2):461--464, 1978.

\bibitem{Kass1995}
Robert~E Kass and Adrian~E Raftery.
 {Bayes Factors}.
 {\em Journal of the American Statistical Association}, 90:773--795,
  1995.

\end{thebibliography}

\end{document}